\documentclass[12pt]{article}
\usepackage{amsmath}
\usepackage{amssymb}
\usepackage{graphicx}
\usepackage{array}
\usepackage{ragged2e} 
\usepackage{setspace}
\usepackage{multicol}
\usepackage{longtable}
\usepackage{geometry}
\graphicspath{{C:/Users/user/Desktop/}}

\usepackage{inputenc}

\begin{document}
	
	\Large
	\begin{center}
		
 \textbf{Birkhoff's Theorem and Lie Symmetry Analysis}

	\large

\vspace{5mm}

A. Mukherjee\footnote{avijit00@gmail.com}\\
\textit{Relativity and Cosmology Research Centre, Department of Physics}\\
\textit{Jadavpur University, India.}\\

and

Subham B. Roy\footnote{sroy3@albany.edu}\\
\textit{Department of Physics}\\
\textit{University at Albany, State University of New York}\\
 
\vspace{3mm}

\textit{Formerly at Relativity and Cosmology Research Centre, Department of Physics, Jadavpur University, India}\\

\vspace{5mm}

\textbf{Abstract}\\

\normalsize

\vspace{5mm}

\justify

 Three dimensional space is said to be spherically symmetric if it admits SO(3) as the group of isometries. Under this symmetry condition, the Einstein's Field equations for vacuum, yields the Schwarzschild Metric as the unique solution, which essentially is the statement of the well known Birkhoff's Theorem. Geometrically speaking this theorem claims that the pseudo - Riemanian space-times provide more isometries than expected from the original metric holonomy/ansatz. In this paper we use the method of Lie Symmetry Analysis to analyze the Einstein's Vacuum Field Equations so as to obtain the Symmetry Generators of the corresponding Differential Equation. Additionally, applying the Noether Point Symmetry method we have  obtained the conserved quantities corresponding to the generators of the Schwarzschild Lagrangian and paving way to reformulate the Birkhoff's Theorem from a different approach.\\
 
\justify
 
 \vspace{10mm}


\vspace{15mm}

\end{center}

\justify

\section{INTRODUCTION}

 \normalsize 
 
\vspace{1mm}

Einstein's Field equations are the most fundamental equation in the realm of General Relativity. This equation can be loosely summarized as the link up between the matter content and the geometry of space-time. In a more qualitative manner, the field equations explains how the metric (of the space-time involved) respond to energy and momentum. Due to the non-linearity and indeterminacy of the equations it is very hard to obtain the solution of the field equations. All the solutions that are available in literature are carried out under simplifying assumptions. The most well known solution of Einstein's equations is obtained under the assumption of spherical symmetry, is known as the Schwarzschild solution.

\vspace{3mm}

 Lie Symmetry analysis of differential equations provides us a rudimentary yet very powerful machinery to derive the conservation laws of corresponding system represented by the (ordinary or partial) differential equations. The invariance of the differential equations under transformation of both dependent and independent variables involved essentially leads to the idea of their symmetry analysis. This transformation forms a local group of point transformation establishing diffeomorphism on the space of independent and dependent variables, mapping the solutions of the specific differential equation to the other solutions.
 
 \vspace{3mm}
 
 In this paper we use the method of Lie Symmetry Analysis on Einstein's field equations. There after we use the Noether's theorem, which reveals the inner relation between the involved symmetries and the conserved quantities of dynamical system to reiterate the well known Birkhoff's theorem.
 
 \vspace{3mm}

The paper is organized in the following manner. In Section 2, we quickly recapitulate the Birkhoff's theorem and discuss about the Lie Algebra of the Killing vectors in a spherically symmetric space-time. Section 3, sheds some light upon the recent literatures that deals with Birkhoff's theorem and it's modern treatments to different dimensions. In the section followed by we introduce the most important tool implemented for our work that is the Lie groups of transformations and the Prolongation Theory of differential equations. In simple language, the term prolongation means becoming longer. Here the involved system certainly does not become longer but the space of the dependent variables does. Basically our requirement is of a differential equation not only representing the dependent and independent variables but also the appearing partial derivatives. So, we prolong the space of dependent variables using their partial derivatives. In section 5, we take the Einstein's vacuum field equations and obtain the maximal symmetry generator, using the method of Lie Symmetry analysis. Entire section 6 comprises of Noether's theorem and it's modification for the first order prolongations. This theorem has been implemented for the case of Schwarzschild lagrangian. This section concludes with recovering Birkhoff's theorem from the infinitesimal symmetry generators obtained via the analysis of Schwarzschild Lagrangian.\\

\vspace{15mm}

		\section{BIRKHOFF'S THEOREM}
		
	\vspace{1mm}
	
	\normalsize

	A metric describing the space-time is obtained when Einstein's field equations are solved under simplifying assumptions. The Birkhoff's theorem is a fundamental theorem regarding the solution of the field equations. It states that, any spherically symmetric solution of Einstein's vacuum field equations must be static and asymptotically flat. This statement leads to the fact that Schwarzschild's exterior (i.e. space-time outside of a spherically non-rotating gravitational object) solution is the most general spherically symmetric solution of Einstein's vacuum field equations with zero cosmological constant. The solution for non-zero cosmological constant is provided by the Schwarzschild-deSitter metric. The Einstein's field equations are given by,

$$ R_{ik} - \frac{1}{2}g_{ik} R = T_{ik}$$

Now for of vacuum, the energy-momentum tensor vanishes i.e. $T_{ik} = 0$

$$ R_{ik} - \frac{1}{2}g_{ik}R = 0 $$

where $g_{ik}$ is the metric of the space-time involved and  $R_{ik}$ and $R$ being the Ricci tensor and Ricci scalar respectively. The solution of this equation for a space-time outside a spherical source or gravitating body is given as,

$$ds^2 = \bigg(1 - \frac{2GM}{c^2 r}\bigg)dt^2 - \bigg(1 - \frac{2GM}{c^2 r}\bigg)^{-1} dr^2 - r^2(d\theta^2 + \sin^2\theta d\phi^2)$$
 
Where $(r,\theta,\phi)$ are the spherical coordinates used and $M$ is the mass of the (spherical) gravitating object. Redefining $ m = \frac{GM}{c^2}$ we get,

$$ ds^2 = \bigg(1 - \frac{2m}{r}\bigg)dt^2 - \bigg(1 - \frac{2m}{r}\bigg)^{-1} dr^2 - r^2(d\theta^2 + \sin^2\theta d\phi^2)$$
 
\vspace{3mm}

The theorem also conveys that the solutions of Einstein's vacuum field equations possess hypersurface orthogonal Killing vectors i.e. the metric obtained as solution of field equations is static\footnote{The description of Birkhoff theorem with Einstein's vacuum solutions having static Schwarzschild metric as solution can be a bit of misleading as well since in the region $0<r<2M$, the time coordinate t fails to remain time like, see Hawking and Ellis Appendix.}. Additionally the theorem also tells us the metric is independent of the changes in the matter distribution, which are the sources  of the gravitational field, provided the spherical symmetry is preserved. 

\vspace{3mm}

The presence of SO(3) group as a group of isometries renders the metric to be spherically symmetric. The existence of Killing vectors corresponding to respective symmetries characterizes the symmetries of space-time or metric. The existing Killing vectors of any given space-time are coordinate independent quantities. None the less by labelling the space-time in specific coordinates, we can formulate the Killing vectors in the chosen coordinate. For example, consider a $S^2$ (having SO(3)) with the Killing vectors labelled as X,Y,Z which in polar coordinates are given as follows, 

$$ X = \partial_\phi$$
$$ Y = (\cos\phi) \partial_\theta - (\cot\theta \sin\phi) \partial_\phi$$
$$ Z = -(\sin\phi) \partial_\theta - (\cot\theta \cos\phi) \partial_\phi$$

A quick evaluation of the closed commutators of these Killing vectors,

\pagebreak

$$ [X,Y] = Z$$
$$ [Y,Z] = X$$
$$ [Z,X] = Y$$

gives us the Lie algebra of SO(3) group.

\vspace{3mm}

Now it is observed that along with these three Killing vectors the Schwarzschild solution does allow one extra time-like Killing vector; 3 for spherical symmetry and 1 for time translation (only in the limit $r>2M$, beyond the Schwarzschild radius the Killing vector remain hypersurface - orthogonal i.e beyond the Schwarzschild radius the time-like coordinte becomes space-like)  \cite{31}. Corresponding to every Killing vector, there will be a constant of motion. For a free particle moving along a geodesic, having of equation of motion 

 $$ \ddot{x}^a + \Gamma^a_{bc} \dot{x}^b \dot{x}^c = 0$$
 
  if $K^\nu$ is a Killing vector then $K^\nu \frac{dx^\nu}{d\lambda}$ remains conserved. The explicit expressions for the four conserved quantities associated with the Killing vectors can easily be written. Summarizing we can conclude that any spherically symmetric solution of the vacuum field equations allow us a fourth/extra Killing vector (we started with SO(3)).

\vspace{3mm}
 
 However, this is one of the many aspects of Birkhoff theorem. There are many different approaches available to this theorem. Such as, physically, it implies that a spherically symmetric star undergoing strict radial pulsations cannot propagate any disturbances into the surrounding space, indicating that a radially oscillating star has a static gravitational field \cite{29}\cite{30}. This paper is based on the differential geometric aspect of Birkhoff's theorem which is stated as, every member of the family of the pseudo-Riemannian space-time has more isometries than expected from the original metric ansatz/holonomy (we have described briefly about this following the pedagogical technique of using the Killing vectors).
  
\vspace{15mm}

	\section{GENERAL FORMULATIONS OF BIRKHOFF'S THEOREM}
	
	\vspace{1mm}
	
	\normalsize
	
	The Birkhoff's theorem concerning Einstein's General theory of Relativity was first proposed and discussed by G.D.Birkhoff (1923) in his paper cited as follows, \textit{The field outside the spherical distribution of matter is static whether or not the matter is in a static or in a variable state...thus the Schwarzschild solution is essentially the most general solution of the field equations with spherical symmetry}\cite{27}.
		
	Over the years Birkhoff's theorem has been addressed in many different ways. Upon stating the fact that the theorem relies on the existence of a 3-parameter group of global isometries with 2-dimensional non-null orbits and of an additional Killing vector associated with a $G_4$ group of motion, H. Goenner \cite{8} had put forward a generalized and geometric version of the Birkhoff's theorem. H.J.Schmidt on the other hand provided a complete covariant proof of Birkhoff's theorem by showing that the origin of Birkhoff's theorem rests on the property that compared to all other dimensions k, it holds for k=2. Other different approaches to this theorem are very lucidly described in his work \cite{10}\cite{7}.
	
	\vspace{3mm}

\cite{32}\cite{33} These two papers describe the 5-dimensional case related to Birkhoff's theorem. Other generalizations such as, extending this theorem to fourth order gravity can be found in the works of P. Havas \cite{34}. The relation of Birkhoff's theorem with 2-dimensional space-time can be found in \cite{35}\cite{36}.

\vspace{3mm}

The most evolved phrasing of this theorem in connection with the conformally reducible metrics can be found in the works of Bona \cite{37}. The author's work focuses on the different kinds of space-time to which the theorem is applicable. It is based on the fact that these space-times are conformal to the direct product of two 2-dimensional manifold. Other generalization of this theorem for higher dimension was done by K.A. Bronnikov and V.N.Melnikov \cite{38} where they have discussed about the validity conditions for the extended Birkhoff's theorem in multidimensional gravity with no resctrictions on space-time dimensionality. An elucidating discussion on the relation between manifold dimensionality and the existance of Birkhoff like theorems has been done by H.J.Schmidt \cite{10}\cite{7}. G.F.R. Ellis and R. Goswami have investigated the possibility of extending the Birkhoff's theorem by analysing whether the theorem holds approximately for an approximate spherical vacuum solution and also for an almost vacuum like configuration \cite{39} \cite{24}.

\vspace{15mm}

	\section{LIE GROUPS OF TRANSFORMATIONS}
	
			\vspace{1mm}
	
	\normalsize
	
	The method of change of variables involed while analysing differential equations is a go-to tool used by physicists and mathematicians alike. Let us illustrate this method by a 2 dimensional case. For example let $x$ and $y$ be the set of variables for a given space. After transformation of the variables we obtain,
	
		$$\hspace{10mm} x \rightarrow x'  = x'(x,y) \hspace{10mm} $$
	$$ \hspace{10mm} y \rightarrow y' = y'(x,y)  \hspace{10mm} $$
	
	where $x'$ and $y'$ are new set of variables involved. This is an example of a Point Transformation which maps points $(x,y)$ into $(x',y')$.
	
	As we are more delved into symmetry properties therefore we will be much more interested in the transformation that also involves parameters.
	
	\vspace{10mm}
	
	\subsection{Groups of Transformation}
	
	\vspace{1mm}

	\normalsize
	
	Let us consider a domain $D \subset \mathbb{R}^N$ (coordinate space) and another space $ S \subset \mathbb{R}^1$ ($S$ be our parameter space). Now let us choose G to be the set of transformations (map) defined by,\\
	
	$$ G : D \times S \rightarrow D$$ 
	
	Now within this set, choose a particular transformation $Z \in G$ defined by,
	
	$$\hspace{20mm} Z : x \times a \rightarrow x' = Z(x ; a)\hspace{25mm}(1)$$

where $x$ and $x'$ are old and new set of coordinates respectively while $a$ being the parameter of transformation involved.

	\vspace{3mm}
	
	(a) For each value of parameter $a \hspace{1mm} \in \hspace{1mm} S$, the transformations are bijective.
	
	\vspace{3mm}
	
	(b) S with the law of composition $\mu$ is a group with identity e.
	
\vspace{3mm}
	
	(c) $Z(x\hspace{0.5mm};\hspace{0.5mm}e) = {x}\hspace{2mm};\hspace{2mm} \forall \hspace{1mm} x \hspace{1mm} \in \hspace{1mm} D.$
	
	\vspace{3mm}
	
    (d) $Z\big(Z(x\hspace{0.5mm};\hspace{0.5mm}a)\hspace{0.5mm};\hspace{0.5mm}b\big) = Z\big(x;\hspace{0.5mm}\mu(a\hspace{0.5mm},\hspace{0.5mm}b)\big) \hspace{1mm};\hspace{1mm}\forall \hspace{1mm}x\hspace{1mm} \in \hspace{1mm} D\hspace{1mm}$ and \hspace{1mm} $\forall\hspace{1mm} a,b\hspace{1mm} \in \hspace{1mm}S.$

	\vspace{10mm}

	\subsection{Lie Groups of Transformation}
	
	\vspace{1mm}
	
	\normalsize
	Furthermore satisfying the axioms that are mentioned above if the following properties,

	\vspace{3mm}
	
	(a) involved parameter $a$ is continuous.
	
	\vspace{3mm}
	
	(b) Z is $C^\infty$ \hspace{0.5mm}; \hspace{0.5mm} $\forall \hspace{1mm} x \in D$ and an analytic function of $a \in S$.
	
	\vspace{3mm}
	
are also satisfied then these axioms together with the ones stated in section 4.1 elevates the group of transformation into the class of Lie groups of transformations.
	
	Expanding equation (1) in Taylor series about a=0,
	
	$$ \hspace{5mm} x' = x + a\frac{\partial Z(x\hspace{0.5mm};\hspace{0.5mm}a)}{\partial a}\bigg|_{a=0} + O(a^2)\hspace{10mm}(2)$$
	
	and defining\footnote{often referred as first fundamental theorem of Lie.},
	
	$$\hspace{10mm} \xi (x) = \frac{\partial Z(x\hspace{0.5mm};\hspace{0.5mm}a)}{\partial a}\bigg|_{a=0}  \hspace{10mm}  $$
	
	we found,
	
	$$\hspace{10mm} x' = x + a\xi(x) \hspace{5mm}(3)$$
	
	where $\xi(x)$ is often termed as auxilliary function of the transformation involved.
	
	\vspace{3mm}
	
	To shed some more light on the physical aspect of the theory we use the following one parameter transformation group.
	
	$$\hspace{10mm} x \rightarrow x' = f(x\hspace{0.5mm},\hspace{0.5mm}\mu\hspace{0.5mm},\hspace{0.5mm}a)\hspace{10mm}(4.1)$$
	$$\hspace{10mm} \mu \rightarrow \mu' = \phi(x\hspace{0.5mm},\hspace{0.5mm}\mu\hspace{0.5mm},\hspace{0.5mm}a)\hspace{10mm}(4.2)$$
	
	where $f$ and $\phi$ being transformation maps.\\ 
	
	With our definition,
	
	$$\hspace{10mm} x'|_{a=0} = f(x\hspace{0.5mm},\hspace{0.5mm}\mu\hspace{0.5mm},\hspace{0.5mm}a) = x\hspace{10mm}(5.1)$$
	$$\hspace{10mm} \mu'|_{a=0} = \phi(x\hspace{0.5mm},\hspace{0.5mm}\mu\hspace{0.5mm},\hspace{0.5mm}a) = \mu\hspace{10mm}(5.2)$$
	
	\vspace{3mm}
	
	These properties ensures the above being a one parameter group of point transformation\cite{4}. A simple example of a one parameter group is given by the rotations
	
	$$ x' = x (\cos a) - \mu (\sin a)$$
	$$ \mu' = x (\sin a) + \mu (\cos a)$$
	
	where a is the involved parameter. On contrary, the reflection
	
	$$ x' = -x$$
	$$ \mu' = -\mu$$
	
	is an useful point transformation which does not constitute a one parameter group. 
	
	\vspace{3mm}
	 
	 The one parameter group and its action are best pictorially observed as motion in the $x-\mu$ plane.
	 
	  Consider an arbitrary starting point $(x_o,\mu_o)$ in that plane (with the involved parameter $a$ being zero). Varying this parameter shifts the starting point along some curve. Again repeating this procedure for different values of $(x_o,\mu_o)$ we can obtain a bunch of curves where under the action of a group, each curve can be transformed into one another and are collectively referred to as orbits of the group \cite{6}.
	
	\vspace{10mm}

	\subsection{Invariants}
	
	\vspace{1mm}
	
	\normalsize
	Let us assume, an $C^\infty$ fucntion $F(x)$, which is said to be an Invariant of the Lie group of Transformation if and only if for any group of transformation the condition,
	
	$$\hspace{10mm}F(x') = F(x) \hspace{10mm}$$
	
	(where $x$ and $x'$ are old and new set of coordinates respectively) holds.
	
	 Infinitesimal generator of the group can be easily exploited to characterize the invariance of a function as is illustrated by the following theorem.
	 
	\vspace{3mm}
	
F(x) is invariant under a coordinate transformation, $x' = Z(x\hspace{0.5mm};\hspace{0.5mm}a)$, if and only if,
	
	$$\hspace{10mm} XF(x) = 0\hspace{10mm} $$
	 
	 where X is called the infinitesimal generator of the transformation. It is given as,
	 
	 $$ \hspace{10mm}X = \xi_1(x)\frac{\partial}{\partial x_1} + \xi_2(x)\frac{\partial}{\partial x_2}+.....  + \xi_N(x)\frac{\partial}{\partial x_N} \hspace{5mm} $$
	 
	 where $\xi_1$, $\xi_2$,.....,$\xi_N$ are set of generalized auxilliary functions for the N dimensional case.
	 
	 \vspace{10mm}
	
	\subsection{Lie Groups of differential equations : Prolongation Theory}
	
	\vspace{1mm}
	
	\normalsize
	
	Consider Lie groups of transformations associated to any given type of differential equation (say F) involving n independent variables $(x^1, x^2, ....., x^n)$  $\in$  $\mathbb{R}^n$ and m dependent variables $(u^1, u^2, ....., u^m)$  $\in$ \hspace{1mm} $\mathbb{R}^m$ and $\mathbb{R}^{m+n}$ be the space of all variables $(x,u)$.
	
	\vspace{3mm}
	
Let us consider the transformations,
	
	$$\hspace{10mm} x'^i = Z^i(x, u, a)\hspace{10mm}(6.1)$$
	$$ \hspace{10mm}u'^k = U^k(x, u, a)\hspace{10mm}(6.2)$$

	This transformation acts on the space $\mathbb{R}^{n+m}$ of variables ($x,u$).
	
	\vspace{3mm}
	
	Let $p^k_i$ denote the derivative of dependent variable $u$ with respect to the independent variable $x$,
	
	$$p^k_i = \frac{\partial u^k}{\partial x^i}\hspace{2mm}; (i=1,2,.....,n \hspace{0.5mm};\hspace{0.5mm}k=1,2,.....,m)$$
	
	\vspace{3mm}
	
The transformation of derivatives with the formula (8) leads us to an extension of the Lie group of transformation which are called \textbf{Prolongations}. This extended group acts on the space with variables ($x^i$, $u^k$, $p^k_i$) rather than ($x^i$, $u^k$), so the $k^{th}$ prolonged space will have ($x^i$ , $u^k$, $p^k_i$, ....., $p^k_j$) as variables, often called the jet space.

\vspace{5mm}

	\subsubsection{First Prolongation:} 
	
	\vspace{1mm}
	
	The infinitesimal generator for the extended group $X^{(1)}$, after first prolongation is given by\cite{2}\cite{4}\cite{6},
	
	$$\hspace{10mm} X^{(1)} = X + \zeta^k_i \frac{\partial}{\partial \phi^k_i}\hspace{10mm}(7)$$

	Where X is the generator of the Lie group of transformation of variable ($x^i$, $u^k$). Given as,
	
	$$ \hspace{10mm}X = \xi^i \frac{\partial}{\partial x^i} + \eta^k \frac{\partial}{\partial u^k}\hspace{10mm}(8)$$

	where,
	
	$$\zeta^i_k = D_i(\eta^k) - p^k_j D_i(\xi^j)\hspace{2mm};(i=1,2,.....,n\hspace{0.5mm}; k=1,2,.....,m)\hspace{2mm}(9)$$
	
	and 
	
	$$ \hspace{10mm}D_i = \frac{\partial}{\partial x^i} + p^k_i \frac{\partial}{\partial u^k}\hspace{30mm}(10)$$

	The operator $D_i$ is the Lie derivative operator.
	
\vspace{5mm}

	\subsubsection{Second Prolongation:} Following the same methods the generator of the extended group $X^{(2)}$, can also be obtained as,
	
	$$ X^{(2)} = X^{(1)} + \sigma^k_{ij} \frac{\partial}{\partial r^k_{ij}}\hspace{5mm}(11)$$
	
	where $r^k_{ij}$ is defined as,
	
	$$r^k_{ij} = \frac{\partial p^k_i}{\partial u_j}\hspace{2mm} ; (k = 1,2,.....,m\hspace{0.5mm}; i,j = 1,2,.....,n)\hspace{2mm}(12)$$

	\vspace{3mm}
	
	and
	
	$$\hspace{12mm}\sigma^k_{ij} = \bar{D}_i (\zeta^k_i) - r^k_{ij} \bar{D}_i (\xi^t)\hspace{20mm}(13)$$	
	$$ \hspace{10mm}\bar{D}_i = \frac{\partial}{\partial x^i} + p^k_i \frac{\partial}{\partial u^k} + r^k_{ij} \frac{\partial}{\partial p^k_j}\hspace{15mm}(14)$$

	\vspace{3mm}
	
	Proceeding in the aforesaid manner, the $k^{th}$ order prolongation can also be found.  Taking into account the cumbersome calculations and our limited need upto second order prolongations for our purpose, we leave the calculations of $k^{th}$ order prolongation for the interested readers.
	
	\vspace{3mm}
	
	We are predisposed towards the symmetry properties of the differential equations, indicating that any transformation T $\in$ G, when applied to any solution of the associated differential equation, maps itself into another solution of the same differential equation, that is to say the $i^{th}$ order differential equation
	
	$$\hspace{10mm} F^{(i)}(x,u,.....)=0\hspace{15mm}(15)$$
		
	remains invarient under any transformation belonging to group G. If $X = \xi^i \frac{\partial}{\partial x^i} + \eta^k \frac{\partial}{\partial u^k}$ be the infinitesimal generator corresponding to equation (6) and if the operator of $k^{th}$ prolongation is denoted by $X^{(k)}$ then equation (6) holds for equation (15), the system of PDE, if and only if
	
	$$ \hspace{10mm} X^{(k)} F^{(i)}\bigg|_{F=0} = 0\hspace{22mm}(16)$$
	
	\vspace{3mm}

	Equation (16) is often referred as the Symmetry Condition or Condition of Invariance.
	
	\vspace{15mm}

		\section{LIE SYMMETRY ANALYSIS OF EINSTEIN'S VACUUM FIELD EQUATIONS:}
		
		\vspace{1mm}
		
		\normalsize
The Einstein's vacuum field equations as we know,

$$\hspace{10mm}R_{ik} - \frac{1}{2}g_{ik}R = 0\hspace{15mm}(17.1)$$
$$\hspace{10mm} R_{ik}\bigg|_{vacuum} = 0\hspace{15mm}(17.2)$$

The general covariance of this equation can be interpreted as the notion of a group of generalized motion. 

The application of Lie symmetry analysis of Einstein's empty space field equations can give rise to the maximal group of the involved point transformations.

\vspace{3mm}

To extend the theory of Lie symmetry analysis to Einstein's equation, we consider a n-dimensional manifold M and along with it, we define a metric tensor $g_{ij}$ for the space-time involved (Riemannian). It is convenient to consider the transformation of the coordinates of space-time rather than the function $g_{ij}$ itself. It is pretty clear that the transformations of the type,

$$ \hspace{10mm} x^i \rightarrow x'^i =  f^i(x,a)\hspace{2mm}; \hspace{2mm} (i=1,2,.....,n)\hspace{10mm} $$ 

where $a$ is the involved parameter. We can write,

$$\hspace{5mm} g_{ij}(x) \rightarrow g'_{ij}(x') = g_{kl}\frac{\partial f^k(x,a)}{\partial x^i}\frac{\partial f^l(x,a)}{\partial x^j}\hspace{10mm}(18)$$

\vspace{3mm}

The generator of the group G,

$$\hspace{20mm} X = \xi^i (x) \frac{\partial}{\partial x^i}\hspace{25mm}(19)$$

where $\xi^i = \frac{\partial f^i}{\partial a}\bigg|_{a=0}\hspace{2mm};\hspace{2mm}(i=1,2,.....,n)$ is the auxilliary function involved.

\vspace{3mm}

Now the generator of the extended group $\bar{G}$,

$$\bar{X} = \xi^i \frac{\partial}{\partial x^i} + \eta_{ij}\frac{\partial}{\partial g_{ij}}\hspace{0.5mm} ; \hspace{0.5mm}\eta_{ij} = \frac{\partial g'_{ij}}{\partial a}\bigg|_{a=0} \hspace{10mm}(20)$$

The preceeding equation along with the fact that when a=0, $g'_{ij}(x) = g_{ij}(x)$ gives us,

$$\eta_{ij} = -\bigg(g_{ik}\frac{\partial \xi^k}{\partial x^j} + g_{kj}\frac{\partial \xi^k}{\partial x^i} \bigg)\hspace{2mm};(i,j = 1,2,3,.....n)\hspace{5mm}(21)$$

The generator of the extended group $\bar{G}$ is 

$$\bar{X} = \xi^i(x)\frac{\partial}{\partial x^i} - \bigg(g_{ik}\frac{\partial \xi^k}{\partial x^j} + g_{kj} \frac{\partial \xi^k}{\partial x^i} \bigg) \frac{\partial}{\partial g_ij}\hspace{5mm}(22)$$

As it is known very well that the Einstein's vacuum field equations allow all possible transformations of coordinates i.e. a infinite dimensional Lie algebra is obtained from equation (22). However the operator satisfied by equation (22) does not admit the maximal algebra. \footnote{The Einstein's equation allows the transformation, $ x'^i \rightarrow x^i $ and  $g'_{ij} \rightarrow ag_{ij}$ which is indeed a simple transformation with a generator, $$X = g_{ij} \frac{\partial}{\partial g_{ij}}\hspace{10mm}(23)$$ which certainly do not belong to the class of generators specified in equation (22)}

\vspace{3mm}

So assuming a generator of the form,

$$ X = \xi^i(x,g)\frac{\partial}{\partial x^i} + \eta_{ij} (x,g)\frac{\partial}{\partial g_{ij}}\hspace{12mm}(24)$$

along with the invariance condition,

$$\hspace{20mm} X^{(2)} R_{ik} = \Omega^{jl}_{ik} R_{jl} \hspace{20mm}(25)$$

where $\Omega^{jl}_{ik}$ are undeterminded coefficients. Segregating the terms with second derivatives of $g_{ij}$ we obtain the Lie determining equations for Einstein's equations resulting in,

$$\hspace{25mm} \xi^i = \xi^i(x)\hspace{55mm}(26.1)$$
$$\hspace{10mm}\eta_{ij} = -\bigg(\frac{\partial \xi^k}{\partial x^i}g_{kj} + \frac{\partial\xi^k}{\partial x^j}g_{ki}\bigg) + ag_{ij}\hspace{35mm}(26.2)$$
$$\hspace{10mm} X = \xi^i(x)\frac{\partial}{\partial x^i} - \bigg(\xi^k\frac{\partial \xi^k}{\partial x^i}g_{kj} + \frac{\partial\xi^k}{\partial x^j}g_{ki}\bigg) + ag_{ij}\frac{\partial}{\partial g_{ij}} \hspace{15mm}(26.3)$$

\vspace{3mm}

Thus it can be seen that the operators given in equation (23) and equation (22) span the maximal Lie algebra admitted by Einstein's empty space field equations.

\vspace{3mm}
 
The generator obtained here is immoderately general for our liking because it produces different sets of infinitesimal generator and different Lie algebras depending upon which space-time or the metric we choose to work on. The generator might give us the gravity group which evidently cannot depend on the metric of choice (just like the Lie algebra of spherically symmetric metric is given as SO(3), independent of the choice of metric.). So drifting away from the far too generalized analysis of Einstein's equation we intend to analyse the geodesic equation of any particular metric (second order is less cumbersome) using the same algorithm.

\vspace{10mm}

	\subsection{Geodesics of Schwarzschild metric:}

\vspace{1mm}

\normalsize
The geodesics of Schwarzschild metric can be obtained by calculating the Christoffel symbol and using the geodesic equation

$$ \hspace{20mm}\ddot{x}^\beta + \Gamma^\beta_{\gamma \mu} \dot{x}^\gamma \dot{x}^\mu = 0 \hspace{20mm}(27)$$

$$\hspace{10mm} \Gamma^\alpha_{\mu \nu} = \frac{1}{2} g^{\alpha l} \big(g_{\mu l, \nu} + g_{\nu l, \mu} - g_{\mu \nu,l} \big)\hspace{10mm}(28)$$

\vspace{3mm}

Using equation (27) and (28) for Schwarzschild metric

$$ds^2 = \bigg(1 - \frac{2m}{r}\bigg) dt^2 - \bigg(1 - \frac{2m}{r}\bigg)^{-1}dr^2 - r^2 d\theta^2 - r^2 \sin^2 \theta d\phi^2 \hspace{10mm}(29)$$

one obtains the geodesic equation for the above metric. The equations are given as,

$$\hspace{10mm}t'' - \frac{a}{r(a-r)}t'(\tau)r'(\tau) = 0\hspace{30mm}(30.1)$$
$$\hspace{-10mm}r''(\tau) + \frac{a}{2r(a-r)} r'^2(\tau) + (a-r) \bigg(\theta'^2 (\tau) + \sin^2\theta \phi'^2(\tau) - \frac{a}{2r^3}t'^2(\tau)\bigg) = 0\hspace{10mm}(30.2)$$
$$ \theta''(\tau) + \frac{2}{r} r'(\tau) \theta'(\tau) - \frac{1}{2} \sin(2\theta) \phi'^2 (\tau) = 0\hspace{15mm}(30.3)$$
$$\phi''(\tau) + \frac{2}{r}r'(\tau)\phi'(\tau) + 2\cot\theta \theta'(\tau) \phi'(\tau) = 0 \hspace{5mm}(30.4)$$

Considering infinitesimal transformation of the form

$$\hspace{10mm}\bar{\tau} = \tau + \epsilon\sigma^\tau \hspace{10mm}  $$
$$\hspace{10mm}\bar{t} = t + \epsilon T^t \hspace{10mm}  $$
$$\hspace{10mm}\bar{r} = r + \epsilon R^r \hspace{10mm} $$
$$ \hspace{10mm}\bar{\theta} = \theta + \epsilon J^\theta \hspace{10mm} $$
$$\hspace{10mm} \bar{\phi} = \phi + \epsilon F^\phi \hspace{10mm}  $$

\vspace{3mm}

and the involved generator X given as,

$$\hspace{10mm}X = \sigma\frac{\partial}{\partial\tau} + T\frac{\partial}{\partial t} + R\frac{\partial}{\partial r} + J\frac{\partial}{\partial\theta} + F\frac{\partial}{\partial \phi} \hspace{10mm} $$

\vspace{3mm}

Running through the usual algorithm of the Lie symmetry analysis of differential equation (reference last chapter) we obtain the vector fields that span the Lie algebra of the space as,

$$ \hspace{30mm}X_1 = \frac{\partial}{\partial \tau}\hspace{25mm}(31.1)$$
$$ \hspace{30mm}X_2 = \frac{\partial}{\partial t}\hspace{25mm}(31.2)$$
$$ \hspace{30mm}X_3 = \frac{\partial}{\partial \phi}\hspace{25mm}(31.3)$$
$$ \hspace{30mm}X_4 = \tau \frac{\partial}{\partial \tau}\hspace{22mm}(31.4)$$	
$$ \hspace{10mm}X_5 = \sin\phi\frac{\partial}{\partial \theta} + \cot\theta\cos\phi\frac{\partial}{\partial \phi}\hspace{10mm}(31.5)$$
$$ \hspace{10mm}X_6 = -\cos\phi\frac{\partial}{\partial \theta} + \cot\theta\sin\phi\frac{\partial}{\partial\phi} \hspace{10mm}(31.6)$$

	\vspace{15mm}

		\section{NOETHER POINT SYMMETRIES}
		
		\normalsize
		
		Emmy Noether\cite{14} proved a very important result that every conservation law pertaining to a system must originate from a corresponding symmetry property of the system. In a more complicated manner, according to Noether's theorem there is an algorithm which relates the constants of the Lagrangian of any given system to its symmetry transformation.
		
		\vspace{3mm}
		
		While dealing with the Lie point symmetry of geodesic equations of a space-time yielding conserved quantities, we realize that they are non-Noether type conserved quantities and hence are redundant to us. The symmetries of the Lagrangian on the other hand being directly connected to the symmetry transformation of a system, gives us conserved quantities of our interest.
		
		\vspace{10mm}
		
		\subsection{Pedagogical Noether's theorem and Prolongations}

		\vspace{1mm}
		
		\normalsize
	Consider a dynamical system with Lagrangian L, where $ L = L(q,\dot{q}, t)$. Using the variational principle to the action we can obtain the Euler-Lagrangian equations written as,
		
		$$ \hspace{20mm}\frac{\partial L}{\partial q^i} - \frac{d}{dt} \bigg(\frac{\partial L}{\partial \dot{q}^i}\bigg) = 0 \hspace{10mm}(32)$$
		
The variational principle involving the action functional can be implemented to formulate the Noether theorem. Here the key constituent is its variations under infinitesimal transformations of the generalized coordinates and time.
		
		$$ \hspace{20mm}q^i \rightarrow \tilde{q}^i = q^i + \epsilon \xi^i (q,\dot{q},t)\hspace{10mm}(33.1)$$
		$$\hspace{20mm} t \rightarrow \tilde{t} = t + \epsilon \eta(q,\dot{q},t) \hspace{10mm}(33.2)$$
		
		\vspace{3mm}
		
		$\epsilon$ is an infinitesimal parameter and $\xi^i$, $\eta$ are analytic functions. The generator of the transformation is given by,	
		$$ \hspace{20mm}X = \eta \frac{\partial}{\partial t} + \xi^i \frac{\partial}{\partial q^i} \hspace{10mm}(34)$$
		
		\vspace{3mm}
		
		The transformation maps the velocities (terms involving first order derivating of the generalized coordinates),
		
		$$\hspace{20mm} \frac{d{\tilde{q}}^i}{d\tilde{t}} = \dot{q}^i + \epsilon (\dot{\xi}^i - \dot{q}^i \dot{\eta}) \hspace{10mm}(35)$$
		
		\vspace{3mm}
		
		Now any function like the lagrangian, depending on the velocities transform like,
		
		$$\hspace{5mm} L(q,\dot{q},t) \rightarrow L(\tilde{q}, \dot{\tilde{q}}, t) = L(q, \dot{q}, t) + \epsilon X^{(1)}(L(q, \dot{q}, t)) + O(\epsilon^2) \hspace{20mm}(36)$$
		
	\vspace{3mm}
	
 $X^{(1)}$ being the first prolongation of the generator X. Given as,
		
		$$ \hspace{10mm} X^{(1)} = X + \big(\dot{\xi}^i - \dot{q}^i \dot{\eta}\big)\frac{\partial}{\partial \dot{q}^i}\hspace{10mm}(37)$$
		
	Due to the infinitesimal shift of the coordinate, the change in action can be put forward as,
		
		$$ \delta A = \int_{\bar{t}_1}^{\bar{t}_2} L\bigg(\tilde{q}, \frac{d\tilde{q}}{d\tilde{t}}, \tilde{t}\bigg) d\tilde{t} - \int_{t_1}^{t_2} L\bigg(q, \frac{dq}{dt}, t\bigg) dt\hspace{10mm}(38)$$
		$$\hspace{20mm} \delta A = \epsilon \int (X^{(1)} L + \dot{\eta}L)dt \hspace{15mm}(39)$$
		
		\vspace{3mm}
		
	We have neglected the higher order terms throughout.	We can conclude that the variation of action is invariant upto divergence term f, if the integrand of equation (39) is total time derivative of some function f(q, $\dot{q}$,t),
		
		$$ \hspace{10mm}X^{(1)} L + \dot{\eta}L = \dot{f} \hspace{10mm}(40)$$
		
	\vspace{3mm}
		
		This is the \textbf{Rund-Trautman} identity which can be used to unearth the carefully hidden symmetries that the Lagrangian fails to exhibit.\\
		
		Now the Rund-Trautman identity is valid for all type of paths $t \rightarrow q(t)$ so it is more logical to replace the dots with total time derivative operator,\\
		
		$$\hspace{10mm}D = \frac{\partial}{\partial t} + \frac{\partial}{\partial q^i} \dot{q}^i + \frac{\partial}{\dot{q}^i} \ddot{q}^i + ....\hspace{10mm}$$
		
		So rewritting the equation (40) as,\\
		
		$$\hspace{10mm}X^{(1)}L + (D \eta)L = (Df)\hspace{10mm}(41)$$
		
		\vspace{3mm}
		
		where the first prolongation operator is given as,\\
		
		$$ \hspace{10mm}X^{(1)} = X + \big(D\xi - \dot{q}(D\eta)\big)\frac{\partial}{\partial \dot{q}} \hspace{10mm} $$
		
		We can solve unknown generator defined in equation (34) using the Rund-Trautman identity, if the Lagrangian of a dynamical system is specified.\\
		
		\vspace{3mm}
		
		If the action functional is invariant under the infinitesimal change of generalized coordinates upto the divergence term f, then the quantity,
		
		$$\hspace{10mm} I = f - L\eta - \frac{\partial L}{\partial \dot{q}^i} \big(\xi^i - \dot{q}^i \eta \big)\hspace{10mm}(42)$$

		is a first integral of the system involved.
		
		\vspace{3mm}
		
		The Lagrangian associated with geodesic equations can be written as,
		
		$$ \hspace{5mm}L(q^i, \dot{q}^i) = g_{\alpha \beta} \dot{q}^\alpha \dot{q}^\beta = g_{\alpha \beta} \frac{dq^\alpha}{d\tau} \frac{dq^\beta}{d\tau} \hspace{10mm}(43)$$
		
	\pagebreak
\subsection{Schwarzschild case}

\normalsize
		
		\vspace{1mm}
		
		For Schwarzschild metric the Lagrangian takes the form,
		
		$$L = \bigg(1 - \frac{2m}{r}\bigg)\dot{t}^2 - \bigg(1-\frac{2m}{r}\bigg)^{-1}\dot{r}^2 - r^2\dot{\theta}^2 - r^2 \sin^2\theta \dot{\phi}^2\hspace{20mm}(44)$$

		(bringing back the conventional terminology instead of using q and t).
		
		\vspace{3mm}
		
		Taking the symmetry generator to be of the standard form,
		
		$$\hspace{4mm} X = \sigma\frac{\partial}{\partial \tau} + T \frac{\partial}{\partial t} + R \frac{\partial}{\partial r} + J \frac{\partial}{\partial \theta} + F \frac{\partial}{\partial \phi} \hspace{10mm}(45)$$

		Substituting this into equation (40) and using the Schwarzschild Lagrangian, we obtain huge cumbersome equations which upon seperation of monomials gives us 16 coupled PDE as listed below.
		
		$$ \hspace{20mm}\sigma = \sigma(s)\hspace{30mm}(46.1)$$		
		$$ R\bigg(\frac{2m}{r^2}\bigg) + 2 \bigg(1-\frac{2m}{r}\bigg)T_t - \bigg(1-\frac{2m}{r}\bigg) \sigma_\tau = 0 \hspace{15mm}(46.2)$$
				$$\hspace{-20mm}R \bigg(\frac{2m}{(r-2m)^2}\bigg) - R_r \bigg(\frac{2r}{(r-2m)}\bigg) +  \bigg(\frac{r}{(r-2m)}\bigg) \sigma_\tau = 0 \hspace{10mm}(46.3)$$
		$$ \hspace{13mm} -2Rr - 2r^2J_\theta + r^2 \sigma_\tau = 0 \hspace{20mm}(46.4)$$
		$$\hspace{-10mm} R(-2r^2 \sin^2 \theta) - 2Jr^2\sin\theta\cos\theta - 2r^2\sin^2\theta F\phi + r^2\sin^2\theta\sigma_\tau\hspace{5mm} = 0\hspace{10mm}(46.5)$$	
		$$ \hspace{10mm}2\bigg(1-\frac{2m}{r}\bigg)T_r - \bigg(\frac{2r}{r-2m}\bigg)R_t = 0 \hspace{10mm}(46.6)$$
			$$ \hspace{10mm} 2\bigg(1 - \frac{2m}{r} \bigg)T_\theta - 2r^2J_t = 0\hspace{20mm}(46.7)$$
		$$ \hspace{10mm}2\bigg(1-\frac{2m}{r}\bigg)T_\phi - 2r^2\sin^2\theta F_t = 0\hspace{10mm}(46.8)$$
		$$\hspace{5mm}-\bigg(\frac{2r}{r-2m}\bigg)R_\phi - 2r^2\sin^2\theta F_r = 0\hspace{12mm}(46.9)$$
		$$\hspace{10mm}-2r^2J_\phi - 2r^2\sin^2\theta F_\theta = 0\hspace{17mm}(46.10)$$
		$$\hspace{10mm}-\bigg(\frac{2r}{r-2m}\bigg)R_\theta - 2r^2J_r = 0\hspace{17mm}(46.11)$$
		$$\hspace{25mm}f_s = 0\hspace{27mm}(46.12)$$
		$$ \hspace{15mm}2\bigg(1-\frac{2m}{r}\bigg)T_\tau = f_t\hspace{15mm}(46.13)$$
		$$\hspace{10mm} -\bigg(\frac{2r}{r-2m}\bigg)R_s = f_r\hspace{10mm}(46.14)$$
		$$\hspace{15mm} -2r^2J_s = f_\theta\hspace{15mm}(46.15)$$
		$$ \hspace{10mm}-2r^2\sin^2\theta F_s = f_\phi \hspace{10mm}(46.16)$$
		
	\vspace{3mm}

		Cumulating the above equations we can obtain the infinitesimal generators of the Noether point symmetry which happens to span the involved Lie algebra as well.
		
		$$\hspace{25mm} X_1 = \frac{\partial}{\partial\tau}\hspace{25mm}(47.1)$$
		$$\hspace{25mm}X_2 = \frac{\partial}{\partial t}\hspace{25mm}(47.2)$$
		$$\hspace{10mm}X_3 = \cos\phi \frac{\partial}{\partial\theta} - \cot\theta \sin\phi\frac{\partial}{\partial\phi}\hspace{10mm}(47.3)$$
		$$\hspace{10mm}X_4 = \sin\phi\frac{\partial}{\partial\theta} + \cot\theta\cos\phi\frac{\partial}{\partial\phi}\hspace{10mm}(47.4)$$
		$$\hspace{25mm}X_5 = \frac{\partial}{\partial\phi}\hspace{25mm}(47.5)$$
		
	\vspace{3mm}
	
	So the obtained infinitesimal Noether Point Symmetry generators involved for the Schwarzschild Lagrangian are nothing but the vectors that span associated Lie Algebra. As we can see from the expression of the infinitesimal symmetry generators, a quick calculation of the closed commutators lead us to,
	
	$$ [X_3,X_5] = X_4$$
	$$ [X_5,X_4] = X_3$$
	$$ [X_4,X_3] = X_5$$
	
	So the generators spanning the Lie algebra follow the SO(3)  structure. This is quite anticipated since the Schwarzschild solution describes a spherically symmetric space-time. A quick comparison between the Killing vectors involved in SO(3) space-time and the infinitesimal generators obtained using Noether Point Symmetry will tell us that the generators are nothing but Killing vectors of the spherically symmetric space-time. But apart from the three generators, satisfying the SO(3) algebra we have also obtained other generators out of which our primary focus will be pertaining to the generator $X_2$.

	\vspace{10mm}
		
	\subsection{Conserved charge and Birkhoff theorem}
	
		\vspace{1mm}
		
		The presence of conserved charges corresponding to the respective Noether Point Symmetries is the next obvious thing ensured by the Noether theorem. We can calculate the conserved quantities using equation (42). But for our purpose, out of the five conserved quantities found we are mostly interested in the conserved charge that corresponds to the particular generator $X_2 = \frac{\partial}{\partial t}$ which is very much intertwined with time translation symmetry.
		
		\vspace{3mm}
		 
		 The conserved quantity corresponding to this generator is,
		
		$$\hspace{10mm} I = \bigg(1-\frac{2m}{r}\bigg)\dot{t} \hspace{10mm}(48)$$
		
		This quantity so obtained is not only a constant of motion but also consistent with the quantity that remains conserved corresponding to a time-like Killing vector. Hence we can conclude that the time translation invariance generator $X_2$ behaves like a timelike Killing vector giving us the extra symmetry that we were looking for since the onset of the problem. 
	
		\vspace{3mm}
	
		We obtained the Noether point symmetries of Schwarzschild geodesic equations along with the infinitesimal generators given in equation (43). Now we can also see that the vectors $X_3$, $X_4$, $X_5$ form SO(3) algebra.
		
		$$\hspace{10mm}[X_3,X_5] = X_4\hspace{10mm}(49.1)$$
		$$\hspace{10mm}[X_5,X_4] = X_3\hspace{10mm}(49.2)$$
		$$\hspace{10mm}[X_4,X_3] = X_5 \hspace{10mm}(49.3)$$
		
	\vspace{3mm}
		
	The Schwarzschild metric which does offer SO(3) algebra and the vector so obtained span the Lie algebra of Noether point symmetries are nothing but the Killing vectors of $S^2$ which we are expected to obtain since we are working with a spherically symmetric metric, see equation (47). The vectors obtained as $X_3$ , $X_4$, $X_5$ are nothing but Killing vectors of SO(3).
		
	\vspace{3mm}
		
		Now along with this group structure of the obtained generators. The conserved quantities corresponding to $\frac{\partial}{\partial t}$ is given by $(1-\frac{2m}{r})\dot{t}$ is in tune with the above mentioned proposition.
		
		We do know that the Killing vector lead us to a constant of motion (for a free particle)\cite{18}. If $K^\mu$ is a Killing vector then,
		
		$$ \hspace{10mm}K_\mu \frac{dx^\mu}{d\lambda} = constant \hspace{10mm}(50)$$
		
	\vspace{3mm}
		
	Hence from equation (48) and (50) we can conclude that the generator $X_2$ we obtained by solving the partial differential equations is actually a Killing vector (time-like as we are in region $r>2m$). The conserved quantity given in the equation (48) is the corresponding constant of the Killing vector.\\
		
		\vspace{3mm}
		
		\hspace{15mm}	We initiated our work with the spherically symmetric metric ansatz exhibiting SO(3) algebra. While calculating the Noether point symmetries of the Schwarzschild Lagrangian we observe the Lie Algebra of SO(3) among the infinitesimal symmetry generators we have obtained which is an expected result as we had started with the spherically symmetric metric (these generators recognised as nothing but the Killing vectors of the SO(3)). In addition to these generators we have also obtained another generator which was identified with the extra Killing vector that shows an additional symmetry than what we had started with. So essentially we have regained the Birkhoff's theorem which states that spherically symmetric vacuum solutions of Einstein's equations allow us a fourth or additional timelike Killing vector (which is what we have obtained).
		
		\vspace{30mm}
	
	\textbf{Acknowledgement:}\\	
	One of the author of this paper, Subham Basanta Roy, would like to thank University Grants Commission (UGC) India, for financial support (JRF Registration Id : 521612) during the course of this work and also The Relativity and Cosmology Centre, Department of Physics, Jadavpur University for allowing to rightfully conduct the research work without any disruptions.

		\pagebreak

			\justify


\begin{thebibliography}{39}
			
			\bibitem{1}
N. H. Ibragimov : ``Elementary Lie Group Analysis and Ordinary Differential Equation''. (Wiley Series in Mathematical Method in practice)
			
			\bibitem{2}
	Peter J. Olver : ``Applications of Lie Groups to Differential Equations" (Graduate Text in Mathematics), Springer, 2nd Edition. (2000)
			
			\bibitem{3}
		G.W. Bluman, S. Kumei : ``Symmetries and Differential equations" (Applied Mathematical Sciences), Springer, $2^{nd}$ Edition. (1996)
					
			\bibitem{4}
L. V. Ovsyannikov : ``Group Analysis of Differential Equations". Academic Press, New York, NY, USA. (1982)
			
			\bibitem{5}
 G.W. Bluman and J.D Cole :``Simlarity Methods for Differential Equations" (Applied Mathematical Sciences), Vol 13, Springer. (1974)
			
			\bibitem{6}
	Hans Stephani and Malcolm McCallum	: ``Differential Equations: Their solution using symmetries". Cambridge University Press. (1989)
			
			\bibitem{7}
	Hans J\"{u}rgen Schmidt : ``The Tetralogy of Birkhoff theorem". Gen. Rel. Grav. vol. 45. (2013)
			
			\bibitem{8}
	H. Goenner : ``Einstein's Tensor and generalization of Birkhoff Theorem". Commun. Math. Phys. 16 34. (1970)
			
			
			\bibitem{9}
	N. P. Konopleva : ``The Birkhoff Theorem and Uniqueness Problem of spherically space-time model in GR". (Preprint Dubna) E4-95-79. (1995)
			
			\bibitem{10}
H. J. Schmidt :	``A New Proof of Birkhoff Theorem". Grav. Cosmol. Vol 3  (1997). 185-190.
			
			\bibitem{11}
	R. Goswami and G. Ellis :	``Almost Birkhoff Theorem in General Relativity". Gen. Rel. Grav. Vol 43 (2011).
			
			\bibitem{12}
Michael T. Samparlis and Andronikos Paliathanesis :	``Lie and Noether Symmetries of Geodesic equation and Collineations", ArXiv: 1101.5769 v1 [gr-qc] 30 (Jan 2011)
						
			\bibitem{13}
	A. H. Bokhari, A. H. Kara, A. R. Kashif and T. D. Zaman : ``Noether Symmetries Versus Killing Vector and Isometries of Spacetimes". International Journal of Theoretical Physics, vol 45, no. 6. (June 2006)
			
			\bibitem{14}
		Emmy Noether : ``Invariant Variations Problems". (1918), Nachrichtea der Akademie der Wissenschaften in G\"{o}ttingen, Mathematisch - Physikalische Klasse, 2, 235-257.
			(English translation in Transport Theory and Statistical Physics, 1(3), 186-207 (1991)).
			
			\bibitem{15}
	A. Z. Petrov : ``Einstein Spaces". Pergamon Press Oxford.  (1969)
			
			\bibitem{16}
	Arthur L. Besse	: ``Einstein Manifolds". New York. (1987)
			
			\bibitem{17}
		Robert M. Wald : ``General Relativity". The University of Chicago Press. (1984)
			
			\bibitem{18}
		Sean M. Carroll : ``An Introduction to General Relativity Spacetime and Geometry". Pearson Publication. (2003)
			
			\bibitem{19}
		R. Adler, M. Bazin and M. Schiffer : ``Introduction to General Relativity". McGraw-Hill, $1^{st}$ edition. (1965)
			
			\bibitem{20}
		 Hans Stephani, Dietrich Krammer, Malcolm McCallum, Eduard Herlt : ``Exact solution to Einstein Field Equations". $2^{nd}$ Edition.  Cambridge University Press.
			
			\bibitem{21}
		Christian Heinicke and Friedrich Hehl :	``Schwarzschild and Kerr Solutions of Einstein field Equations: An Introduction".
			
			\bibitem{22}
		 K. Schwarzschild : ``\"{U}ber das Gravitations - feld eines Massenpunktes nach der Einsteinschen Theorie". (1916)
			
			\bibitem{23}
	Sir Roger Penrose :	``The Road to Reality: A complete Guide to the Laws of the Universe". New York. (2005)
			
			\bibitem{24}
			 G.F.R. Ellis and R. Goswami, Gen. Rel. Grav. 43,2157 (2011), ArXiv: 1101.4520 [gr-qc]
			
			\bibitem{25}
			N. H. Ibragimov : Selected Works. vol (1 - 5).			
			
			\bibitem{26}
		N. H. Ibragimov : ``Transformation Groups Applied to Mathematical Physics".
			
			\bibitem{27}
		 G.D. Birkhoff : ``Relativity and Modern Physics". p.255. Cambridge : Havard University Press. (1923)
			
			\bibitem{28}
			Jebson, J.T : Ark. Mat. Astron. Fys. 15,1 (1921)  ; Alexandrow, W : Ann, Physik 72, 141 (1923)
			
			\bibitem{29}
		Alan Maciel, Morgan Le Dellion, Jose P. Mimoso :	``Revisiting Birkhoff theorem froma  dual null point of view". ArXiv : 1803.11547 v2 [gr-qc] 6 April 2018.
			
			\bibitem{30}
	       G.E.M.P Hobson and A.N. Lasenby : ``General Relativity : An introduction to physicists".
			
			\bibitem{31}
			S.W. Hawking and G.F.R. Ellis : ``Large Scale Structures of Space-Time" (Cambridge Monographs of Mathematical Physics), Cambridge University Press, Revised Edition. (1975)
			
			\bibitem{32}
		W. Rindler : ``Birkhoff's theorem with $\Lambda$-term and Bertrotti-Kasuer space". Phys.Lett.A 245 (1998)363.
			
			\bibitem{33}
		M. Bojowald, H. Kartrup, F. Schramn, T.Strobl :	``Group Theoretical quantisation of phase space $S^1 \times \mathbb{R}^+$ and the mass spectrum of Schwarzschild black hole in D dimension". Phy.Rev D 62 (2000).
			
			\bibitem{34}
	P. Havas :	``Theories of gravitation with higher order field equations". Gen.Relat.Grav.8 (1977) 631.
			
			\bibitem{35}
	H.J.Schmidt :	``Scale invariant gravity in two dimensions". Jour. Math. Phys. 32 (1991) 1562.
			
			\bibitem{36}
	S.Mignemi and H.J.Schmidt :	``Two dimensional higher derivative gravity and conformal transformation". Class.Quant.Grav.12 (1995) 849.
			
			\bibitem{37}
	C.Bona : ``A new proof of the generalized Birkhoff Theorem". Jour. of Math. Phys. 29,1440 (1988)
			
			\bibitem{38}
			K.Bronnikov and V.N.Melnikov, ``The Birkhoff Theorem for Multidimensional Gravity", Gen.Rel.Grav. 27,465 (1995).
			
			\bibitem{39}
			G.F.R. Ellis and R. Goswami, Gen. Rel. Grav. 44,2037 (2012), ArXiv: 1202.0240 gr-qc
		
			
			
			
			
			
			
			
			
			
				
	
	
			
			
		\end{thebibliography}
\end{document}